\documentclass[12pt]{iopart}
\usepackage{graphicx}
\usepackage[T1]{fontenc}
\usepackage[utf8]{inputenc}
\usepackage{textcomp}
\usepackage{gensymb}
\usepackage{bibentry}

\usepackage{iopams}  
\begin{document}

\title{Electronic properties of atomically coherent square PbSe nanocrystal superlattice resolved by STS}

\author{Pierre Capiod}
\address{Debye Institute for Nanomaterials Science, Utrecht University, PO Box 80 000, 3508 TA Utrecht, the Netherlands}
\author{Maaike van der Sluijs}
\address{Debye Institute for Nanomaterials Science, Utrecht University, PO Box 80 000, 3508 TA Utrecht, the Netherlands}
\author{Jeroen de Boer}
\address{Debye Institute for Nanomaterials Science, Utrecht University, PO Box 80 000, 3508 TA Utrecht, the Netherlands}
\author{Christophe Delerue}
\address{Univ. Lille, CNRS, Centrale Lille, Univ. Polytechnique Hauts-de-France, Junia, UMR 8520 - IEMN, F-59000 Lille, France}
\author{Ingmar Swart}
\address{Debye Institute for Nanomaterials Science, Utrecht University, PO Box 80 000, 3508 TA Utrecht, the Netherlands}
\author{Daniel Vanmaekelbergh}
\address{Debye Institute for Nanomaterials Science, Utrecht University, PO Box 80 000, 3508 TA Utrecht, the Netherlands}
\ead{d.vanmaekelbergh@uu.nl}

\vspace{10pt}
\begin{indented}
\item[]October 2020
\end{indented}

\begin{abstract}
Rock-salt lead selenide nanocrystals can be used as building blocks for large scale square superlattices via
two-dimensional assembly of nanocrystals at a liquid-air interface followed by oriented attachment. Here we report
measurements of the local density of states of an atomically coherent superlattice with square geometry made from PbSe
nanocrystals. Controlled annealing of the sample permits the imaging of a clean structure and to reproducibly probe the
band gap and the valence hole and conduction electron states. The measured band gap and peak positions are compared to
the results of optical spectroscopy and atomistic tight-binding calculations of the square superlattice band structure.
In spite of the crystalline connections between nanocrystals that induce significant electronic couplings, the
electronic structure of the superlattices remains very strongly influenced by the effects of disorder and variability.
\end{abstract}

\section*{Introduction}

Two-dimensional semiconductors, often called semiconductor quantum wells, are key materials in opto-electronics.
Moreover, electron (hole) gasses confined to two dimensions have been extensively investigated, which led to landmark
discoveries such as the quantum Hall\cite{vonklitzing1986} and quantum spin Hall effect\cite{bernevig2006,konig2007}. It
has been shown recently that an extra nanoscale periodicity in the plane can have a major impact on the band structure
of the semiconductor\cite{park2008,park2009,kalesaki2014}. The band gap of the semiconductor remains quasi unaltered,
while the shape of the valence and conduction bands is determined by the geometry. A prominent example is the honeycomb
nanoscale geometry; theory on several levels has shown that the highest valence and lowest conduction bands feature
Dirac cones\cite{kalesaki2014}. This offers the possibility of massless electron (hole) excitations in a genuine
semiconductor.  Other examples of a nanoscale geometry are the square and the Lieb geometry, which also have a
lattice-specific band structure. These band structures have recently been confirmed in artificial lattices based on the
electron gas present on a Cu(111)\cite{gomes2012,slot2017,slot2019} surface, and by coupling vacancy states using a
chlorine monolayer on Cu(100)\cite{drost2017}.

The fabrication of two-dimensional semiconductor materials with a specific nanoscale geometry is still in its infancy.
Lithography and etching techniques have been used to craft a honeycomb geometry in GaAs and InGaAs
surfaces\cite{wang2018,nadvornik2012,post2019,franchinavergel2021}. Due to the large unit cell, the bandwidth and
electronic dispersion in these systems is still limited (up to 7 meV for a lattice with a period of 36
nm)\cite{franchinavergel2021}.  In a bottom-up attempt, PbSe semiconductor nanocrystals were used as building blocks to
prepare two-dimensional nanostructured semiconductors\cite{boneschanscher2014}. PbSe nanocrystals are absorbed at an
interface, become crystallographically aligned, and finally attach in an epitaxial way to form atomically coherent
two-dimensional systems. Self-assembly and oriented attachment with the formation of crystallographic necks between the
nanocrystals resulted in lattices with a square array of nanocrystals (and voids) or a hexagonal array of voids, i.e. a
honeycomb geometry. In coherently attached superlattices, the overlap strength and directions of wavefunctions between
nanocrystals in a square or honeycomb geometry should be much larger than for nanocrystal arrays in which the
nanocrystals are still separated by capping molecules. In the latter, no crystalline necks are formed and the quantum
mechanical coupling arise from the interlacing of the ligands at the surface of
nanocrystals\cite{liljeroth2006,overgaag2008}. This approach allows the engineering of material band structures by
controlling the constituent elements of nanocrystals and the structural parameters (nanocrystal sizes, neck sizes,
nanogeometries) of the arrays. These tunable parameters open a completely new class of material to explore. However,
these systems still suffer from inhomogeneities in the nanocrystal building block size and in the coupling between the
nanocrystals (neck sizes, misorientation between two adjacent nanocrystals creating faulty interfaces), partially
clouding the lattice-specific electronic band
structure\cite{peters2018,whitham2016,ueda2020,maier2020,walravens2019,smeaton2021,chu2020,ondry0000,chen2020}.  These
inhomogeneities rise questions on electronic couplings and dispersions, and on surface potential fluctuations that need
to be understood on simple model systems in order to probe the band structure on more complex systems such as the
honeycomb nanogeometry where Dirac cones and flat bands have been predicted\cite{kalesaki2014}.

Here, we investigate the band structure of a PbSe superlattice with square geometry, and  epitaxial neckings between
the nanocrystals using scanning tunneling microscopy and spectroscopy. This method is well-suited for the investigation
of disordered systems as the local geometry and the local electronic density of states can be obtained\cite{swart2016}.
We compare the band gap, valence hole and conduction electron states with the results of atomistic calculations. Our
results indicate a moderate coupling between the nanocrystal sites of the lattice, despite the nanocrystals being
epitaxially connected. We conjecture that significant potential barriers are still present between the nanocrystals due
to inhomogeneities in the connections. We place our findings in the framework of electronic transport studies that have
been reported on these lattices\cite{whitham2016,ueda2020,evers2015}. Finally we make suggestions to improve the
electronic coupling and lower the disorder.  \\

\section*{Experimental}

The synthesis of PbSe nanocrystals (NCs) is adapted from a well established procedure in the
literature\cite{peters2018,alimoradijazi2019,steckel2006,moreels2007}. For the formation of the square superlattice, the
nanocrystals which are dispersed in a non-polar solvent (toluene) are drop-casted on top of a liquid substrate (ethylene
glycol). As described in the literature\cite{evers2013,vanoverbeek2018,geuchies2016}, the PbSe nanocrystals
self-assemble into a 2D square array with the <100> axis pointing upward (perpendicular to the air-liquid interface),
and finally attach via <100>/<100> facet-to-facet connection upon evaporation of the toluene solvent. With our sample,
the evaporation was done at 22°C for 1.5 hours, then the liquid substrate was heated to 38°C for 20 minutes to
strengthen the necks between the nanocrystals.  After formation of the 2D structure, the superlattice is transferred to a
Highly Ordered Pyrolytic Graphite (HOPG) substrate.  This is done by approaching the substrate parallel to the liquid
substrate until a meniscus is formed between the two.  Then, the PbSe super-structure is washed with chloroform and
immersed in a solution of saturated cadmium iodide in methanol for a (partial) ligand exchange procedure.  Iodide and
oleate create a hybrid ligand shell which provides an excellent protection from oxidation\cite{peters2019}. Further
washing with methanol removes the exchanged ligands from the superstructure surface. After these procedures, the sample
was transferred to the vacuum chamber of a STM for cryogenic microscopy and spectroscopy (see below).

\begin{figure}[!htbp]
	\begin{center}
		\includegraphics[width = 140mm]{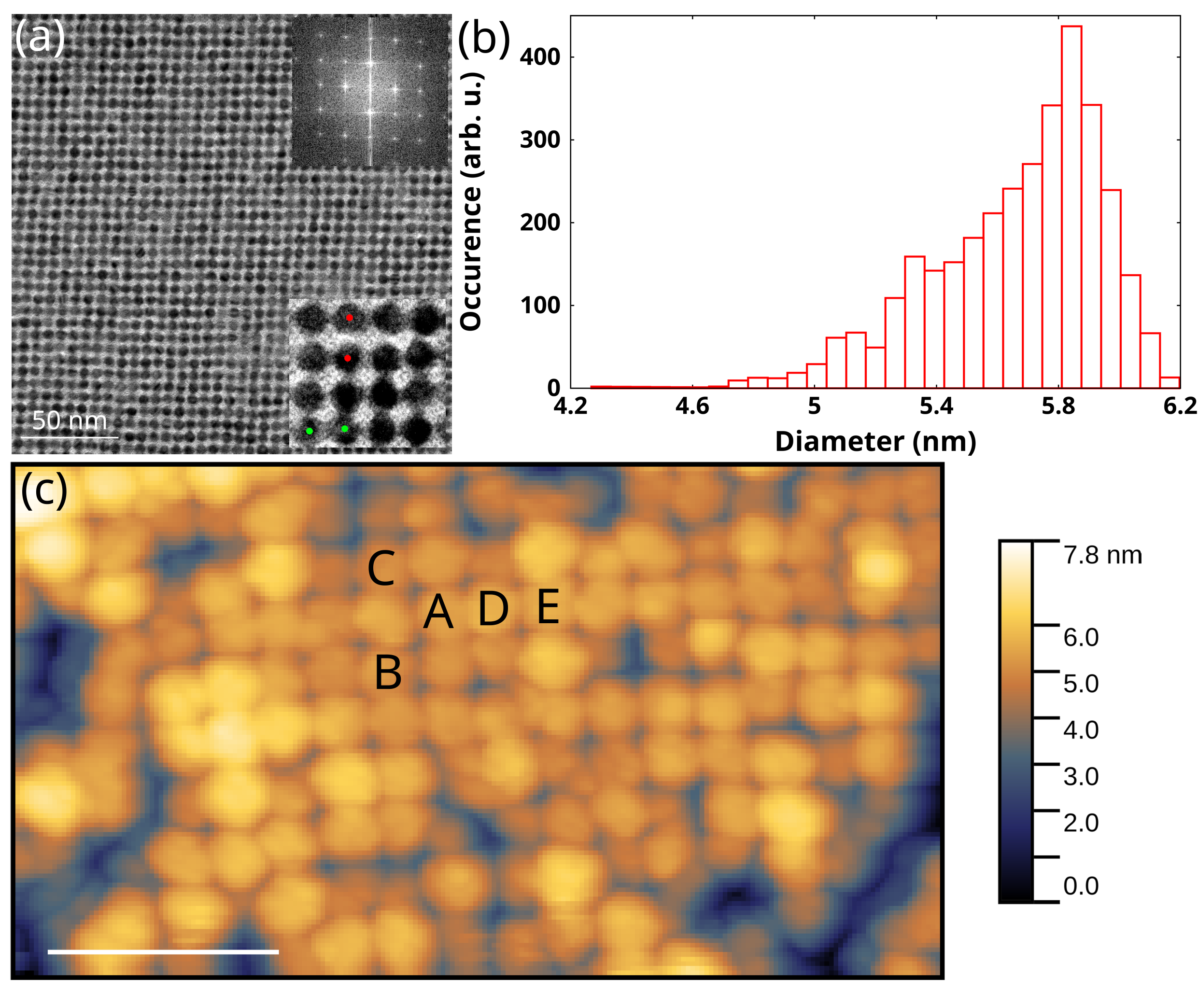}
	\end{center}
        \caption{(a) Transmission Electron Microscope image of the square superlattice, the nanocrystals are in darker
            contrast while the voids in the lattice are in bright. Insets: top right, fast Fourier transform of the
            image presenting a center-to-center distance of 6.4 nm $\pm$ 0.2 nm and bottom right, zoom of the TEM image
            displaying non-existent necking (bridge between the two nanocrystals labeled with red dots) and strong
            necking (bridge between the two nanocrystals labeled with green dots) (b) Nanocrystal size distribution
        measured over 3 TEM images (c) STM image of a PbSe square superlattice ($I$ = 5 pA, $V_{\rm{gap}}$ = 2 V).
    Spectra have been taken on the labeled nanocrystals. The scale bar represents 20 nm.}
	\label{fig:figure1}
\end{figure}

In parallel to the preparation of a STM sample, a part of the same superlattice is characterized by Transmission
Electron Microscopy (TEM), see figure \ref{fig:figure1}(a). The TEM image shows a seemingly regular square lattice, with
a periodic array of voids. The nanocrystal size-distribution extracted from TEM images shows a reversed log-normal
distribution (figure \ref{fig:figure1}(b)). The peak of the size distribution is centered around a diameter of $D$ =
5.84 nm. Fast Fourier transform of the TEM shows a center-to-center distance of 6.40 nm $\pm$ 0.20 nm (top-right inset
of figure \ref{fig:figure1}(a)), the difference is due to formation of elongated necks in the <100>-type directions.
This result is consistent with the observations of Geuchies \textit{et al.}\cite{geuchies2016}. A careful inspection of
the TEM images reveals that the necking varies from nanocrystal to nanocrystal. In the bottom-right inset of figure
\ref{fig:figure1}(a), the necking can range from very pronounced (for nanocrystals marked with green dots) with a large
$d$/$D$ ratio ($d$ being the diameter of the neck,  $D$ the diameter of the nanocrystal) to non-existent (for
nanocrystals marked with red dots), consistent with the literature\cite{whitham2016,evers2015,evers2013}. On the vast
majority of the nanocrystals, the ratio $d$/$D$ to the neighboring nanocrystals is not homogeneous. This non-homogeneity
in the crystalline bridges and nanocrystal sizes is reflected in the band structure (see below for details).  \\

Superlattices prepared by wet chemistry methods form a challenge for cryogenic scanning tunneling microscopy and
spectroscopy\cite{swart2016}.  The presence of ligands and solvent molecules on the superlattice can impede scanning
tunneling microscopy and spectroscopy. Therefore, it is crucial to properly prepare the sample before scanning tunneling
experiments.  We introduced the prepared samples (see above) into the preparation chamber of a LT-STM (Scienta Omicron)
and annealed to a temperature of 140°C for 24 hours in order to remove the excess of ligands, molecules and water at the
surface of the 2D structure. The sample is then placed in the LT-STM head where it is cooled down to 4 K for microscopy
and spectroscopy. A STM image of the square lattice is presented in figure \ref{fig:figure1}(c). It should be remarked
here that stripy artifacts due to ligands are not observed, similar to the results from Ueda \textit{et al.} after
\textit{in-vacuo} passivation of a PbSe quantum dot solid with trimethylaluminum vapor\cite{ueda2020}. We conclude that
the hybrid inorganic/organic shell on our PbSe superlattice, created by the procedure described above, provides a clean
and stable tunneling junction.  However, in certain regions of the sample, blurry lines can still be seen (see S.I.),
which implies that some ligands are loosely bound to the NC surface and can interact with the scanning tip.  By scanning
at a sufficiently low set-point current, it was possible to image the sample with only minor disturbances coming from
the ligands. On the local scale, disorder in the lattice is visible. The added 3rd dimension of the STM image reveals
height differences (from 1 to 2 nm) on the sample surface which cannot be seen in the TEM image. These height
differences together with irregularities in the in-plane directions indicate the degree of structural disorder in the
superlattice.  We assign these imperfections to the size distribution of the building blocks and inhomogeneities in the
epitaxial connections of the nanocrystals occurring in the two (001) directions of the rock-salt crystal
structure\cite{peters2018,walravens2019,li2012}.

The tunneling spectra, presented in figure \ref{fig:figure2}(a), were acquired at constant height by placing the tip
above a single NC site. A PtIr tip has been prepared by indentations into a copper surface prior to inserting the NC
sample\cite{castellanos-gomez2012,Tewari2017}. The feedback loop is disconnected while a variable voltage is applied to
the tunneling junction. The tunneling current $I$ and conductance $dI/dV$ are measured simultaneously. The differential
conductance is obtained with a lock-in amplifier (rms modulation of 10 mV at 271 hz). All spectra were averaged using at
least 20 $dI/dV$ sets of reproducible curves. Five spectra (taken on five different NCs, labeled A, B, C, D and E on
figure \ref{fig:figure1}) are displayed in figure \ref{fig:figure2}(a). Additional spectra taken on different regions
and on a second sample, showing the same electronic features, are shown in S.I. The raw zero-conductivity gap is
determined to be 0.78 eV $\pm$ 0.03 eV by measuring the energetic spacing of the two broad tunneling resonances visible
on both sides of the zero-conductance gap.

In order to accurately estimate the value of the band gap and positions of the peaks, we need to take into account the
polarization energies and the lever arm\cite{franceschetti2000,niquet2002}. The polarization energy is the energy needed
to put an additional charge in the valence and conduction bands, determined by the difference of dielectric constant of
the superstructure and its surrounding.  In the case of isolated PbSe nanocrystals of 5.80 nm with a dielectric constant
$\epsilon_{NC}$ = 227, covered by ligands of a dielectric constant $\epsilon_{out}$ = 3, deposited on a metallic
surface, the polarization energies can be estimated to be around 70 meV\cite{niquet2002}. Since the NCs are in a
superlattice on a conductive surface, the dielectric mismatch is far less important than for isolated nanocrystals due
to the presence of four neighboring nanocrystals of high dielectric constant and the metallic screening.  Hence, the
polarization energy can be neglected for the present case (value around 1 meV).  The lever arm $\eta$ corresponds to the
ratio of the voltage drop in the tip-nanocrystal junction over the entire voltage drop in the tip-nanocrystal-substrate
junction.  Typical values are 0.8 for isolated PbSe nanocrystal of 6 nm\cite{swart2016} and 0.85 for an assembly of 5.30
and 7.30 nm PbSe NCs\cite{liljeroth2006}.  We therefore set the lever arm to 0.85. This means that the single particle
band gap of the superstructure, derived from STS, is 0.66 $\pm$ 0.02 eV (see S.I. for additional sites and sample).
These values are consistent with the values recently reported in the literature on similar PbSe QD
solids.\cite{ueda2020}.  Now, we compare this value with theoretical results. The band structure and density of states
of the PbSe superlattice has been modeled with a tight-binding theory\cite{kalesaki2014,kalesaki2013,delerue2014}. The
model includes broadening of the resonance peaks that describes inter-valley and electron-phonon couplings as main
contributions (see below).  The system is built with spherical nanocrystals and two sizes have been considered $D_1$ =
5.51 nm and $D_2$ = 6.12 nm (respectively, $9 \times a_0$ and $10 \times a_0$ with $a_0$, the lattice constant of PbSe).
The nanocrystals are connected via crystalline cylinders of diameters $d_1$ = 2.76 nm and $d_2$ = 3.06
($d_{1,2}$/$D_{1,2}$ = 0.5, see below for details about this value) in diameter which define the coupling strength
between nanocrystals\cite{kalesaki2014,kalesaki2013}. The tight-binding parameters for PbSe are taken from Ref.
\cite{allan2004}. In the case of the square superlattice, tight-binding calculations give a value of 0.68 eV and 0.62 eV
for the band gap, respectively for $D_1$ and  $D_2$ (curve labeled \textit{TB$_1$} and \textit{TB$_2$} in the figure
\ref{fig:figure2}(a)). The experimental band gap values lie in between the calculated ones, consistent with the particle
size distribution.  \\

%It is important to note that the value of the coupling strength is difficult to estimate from the TEM pictures and from
%the STS only as it is a local average of the coupling with the neighboring nanocrystals. Each connection can be
%different and it is expected to have slight shifts of the on-site energy and different broadenings for each measured
%nanocrystal. Thus, the experimental values correspond to a set of nanocrystal size and coupling strength that can be
%fitted with theoretical calculations such as tight-binding.

%The final values for the calculations presented here have been found after running several calculations and confronting
%the band gap and the bandwidth of the conduction bands against the experimental values.

%While the band gap is on the smaller side due to the size limitation ($10 \times a_0$ = 6.12 nm is closer to measured
%size of the nanocrystals from TEM than $9 \times a_0$ = 5.5 nm), the bandwidth is very similar in for both calculated
%sizes (see figure 2.b). It is clear that the calculated bandwidth is similar to the experimental value when the
%coupling strength $d$/$D$ is close to 0.4.

\section*{Discussion}

\begin{figure}[!htbp]
	\begin{center}
		\includegraphics[width = 140mm]{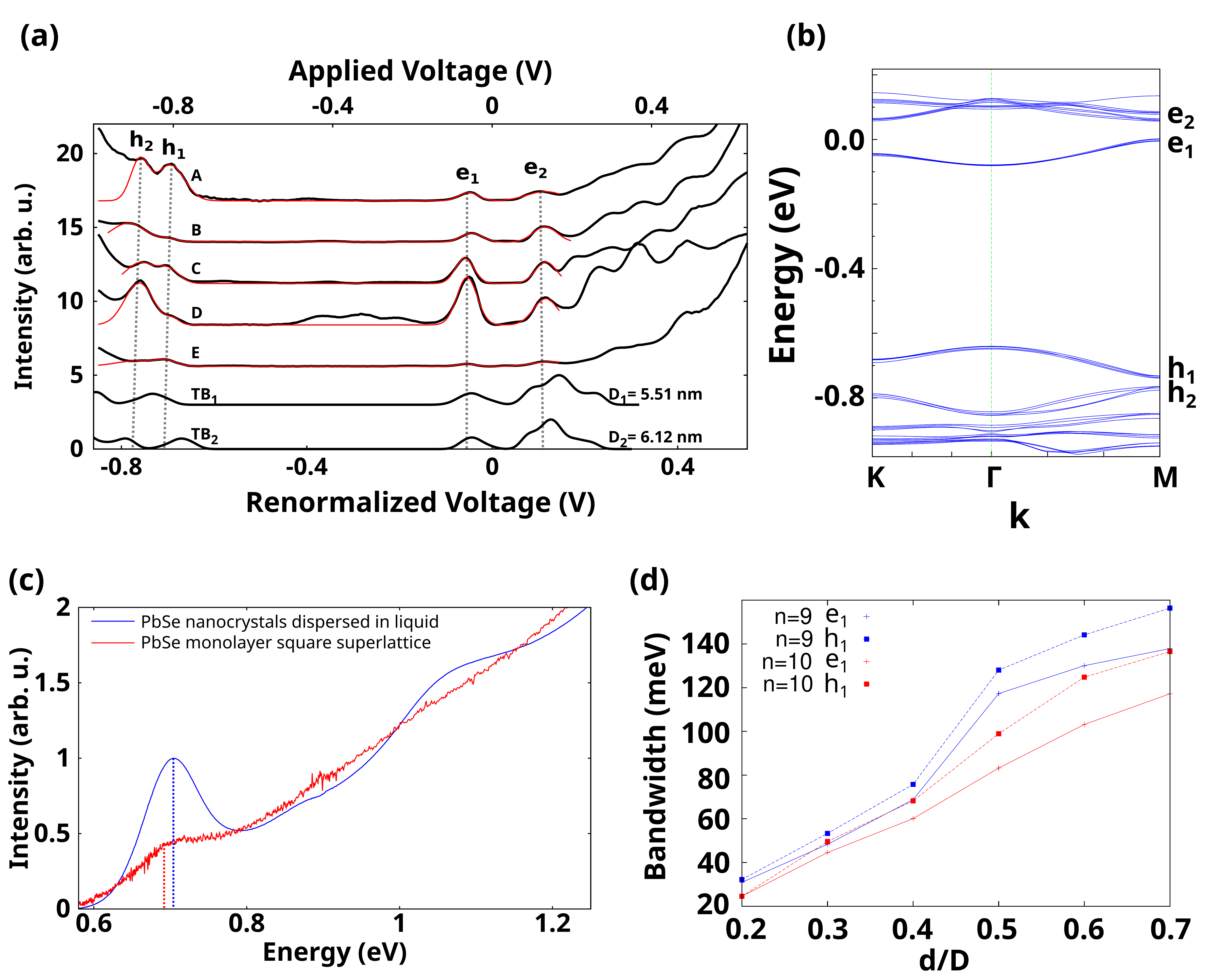}
	\end{center}
    \caption{(a) Tunneling spectroscopy ($I$ = 500 pA) on five different nanocrystals in the superlattice with the
        corresponding Gaussian fits of the peaks in red. A lever-arm $\eta$ = 0.85 has been taken into account to define
        the renormalized voltage. The projected density of states from the tight-binding calculation is shown at the
        bottom. (b) Band structure of the PbSe square superlattice calculated with a unit cell composed of one
        nanocrystal of $D$ = 6.12 nm and projected along the direction K-$\Gamma$-M. (c) Absorption spectra of dispersed
        PbSe nanocrystals in solution and of a monolayer of PbSe superlattice. The red shift indicates a coupling
        between the nanocrystals. The dotted line shows the position of the first level transition (1S$_{\rm{h}}$ to
        1S$_{\rm{e}}$) in both measurements, reflecting the value of the band gap. For the dispersed nanocrystals, the
        value is 0.71 eV and for the superlattice, the band gap is 0.68 eV. Amplitude of red spectrum has been
        renormalized with respect to the blue spectrum to better show the difference in peaks positions. (d) Bandwidth
        of the s-type bands in conduction (e$_{\rm{1}}$) and valence (h$_{\rm{1}}$) bands for the two considered
    nanocrystal sizes $D_1$ and $D_2$}
	\label{fig:figure2}
\end{figure}

In the following, we discuss the resonances observed in the tunneling spectra. We first remark that these resonances
were robust, reproducible (over two different samples, see S.I.) and stable over more than 20 scans. The results
presented in \ref{fig:figure2}(a) are averaged over 20 scans to improve the signal-to-noise ratio. We observe four
resonances (peaks in $dI/dV$) in the positive bias range, reflecting electron conduction states and two resonances in
the negative bias related to the valence hole states. The first peak (labeled e$_{\rm{1}}$ in the figures
\ref{fig:figure2}(a) and \ref{fig:figure2}(b)) of the conduction band, located at -0.05 eV $\pm$ 0.02 eV reflects the
coupling of nanocrystal states with 1S$_{\rm{e}}$ envelope wavefunction. It is situated below the Fermi level (zero
bias), suggesting a heavily doped structure. The composition of the PbSe nanocrystal building blocks is not
stoichiometric\cite{moreels2007}.  There is a variable excess of Pb atoms at the surface of the nanocrystals (due to the
Pb(OA)$_2$ that binds to dangling Se$^{2-}$ at the surface) which causes an unintentional n-doping of the
structure\cite{evers2015,gai2009,oh2013}. The second peak (e$_{\rm{2}}$) situated at 0.11 eV $\pm$ 0.01 eV in the
conduction band is attributed to the 1P$_{\rm{e}}$ envelope.

In the valence band, the first peak (h$_{\rm{1}}$), situated at -0.70 eV $\pm$ 0.01 eV is the 1S$_{\rm{h}}$.  The
second peak h$_{\rm{2}}$ at -0.78 eV $\pm$ 0.01 eV is attributed the 1P$_{\rm{h}}$ group. The relative close values
of the 1S$_{\rm{h}}$ and 1P$_{\rm{h}}$ are due to the different valence band maxima (located at the L, $\Sigma$ and
near the K point of the Brillouin zone) of the PbSe band structure for QDs\cite{an2006}. 

By taking into account an energy shift of -0.51 eV of the calculated electronic levels due to the doping, the conduction
resonances and valence resonances agree well with the atomistic tight-binding calculations for a square PbSe lattice,
see figure \ref{fig:figure2}(a).  Figure \ref{fig:figure2}(b) presents the band structure calculations based on a
building block size of 6.12 nm and a coupling $d$/$D$ = 0.5. The different band folds are clearly visible and labeled as
presented above.

Optical absorption spectroscopy has been conducted in parallel to the STS experiments. Figure \ref{fig:figure2}(c) shows
the absorption spectra for the dispersed nanocrystals in solution (before attachment) and for the square superlattice.
The first excitonic peak for both curves represents the 1S$_{\rm{h}}$ to 1S$_{\rm{e}}$ transition and is a direct
measurement of the band gap\cite{an2006}.  Superlattice formation results in a lowering of the band gap from 0.71 eV to
0.68 eV, which agrees well with the band gap measured by STS and the tight-binding calculations. With the observed
reduction of the QD size, the measured band gap should have increased. However, the decrease of the band gap upon
supercrystal synthesis suggests that there is a formation of a 2D system with creation of crystalline necks between
nanocrystals of the superlattice.  The broadening of the first optical transition results from this quantum mechanical
coupling, but may also reflect local disorder and the effects of light scattering. \\

\begin{figure}[!htbp]
	\begin{center}
        \includegraphics[width = 160mm]{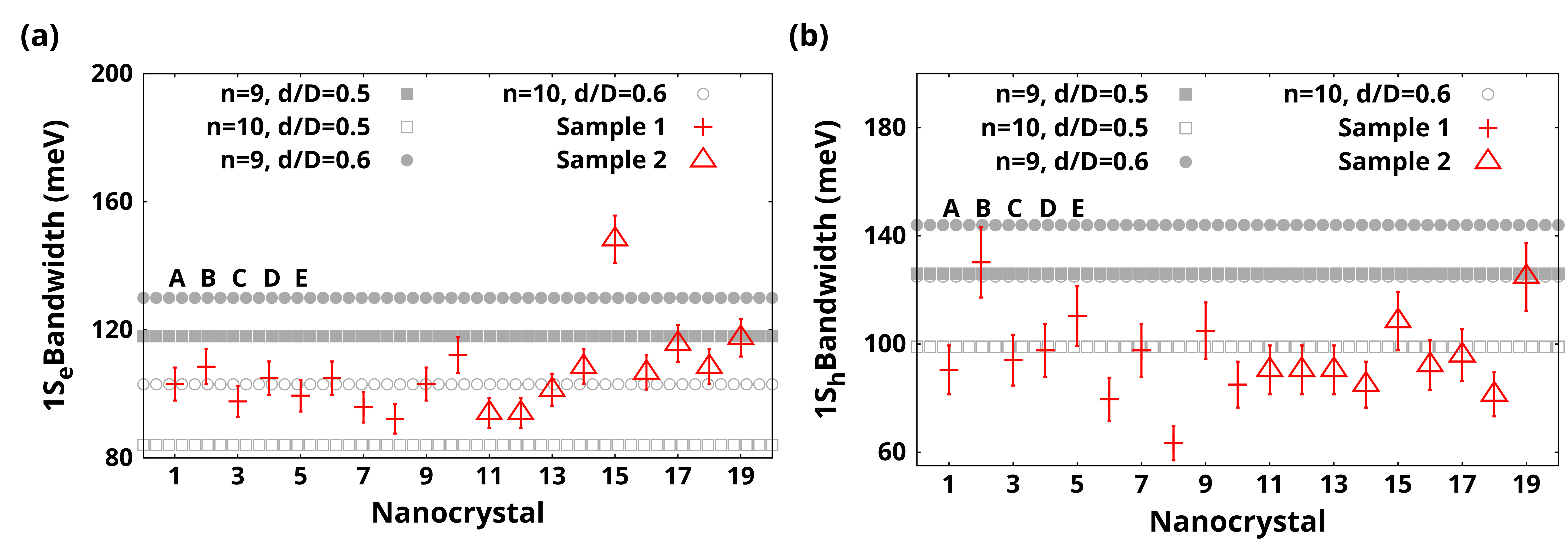}
	\end{center}
    \caption{(a) Bandwidths of the 1S$_{\rm{e}}$ bands in the conduction band for two different samples (crosses for
        the sample 1, triangles for sample 2, see S.I. for more details) compared to calculated dispersions for
    superlattices composed of nanocrystals with diameter of 5.5 nm (filled symbols) and 6.1 nm (open symbols),
respectively n = 9 and n = 10 and for each, two necking sizes $d$/$D$ = 0.5 (circles) and  $d$/$D$ = 0.6 (squares). (b)
Same comparison for the valence band including the bandwidths of the 1S$_{\rm{h}}$ bands. Labels "A" to "E" correspond
to the nanocrystal labelled on figure \ref{fig:figure1}.}
    \label{fig:figure3}
\end{figure}

We now turn to the width of the observed STS peaks, as these provide information about the quantum mechanical coupling
(hopping parameter in a tight-binding point of view) between nanocrystals\cite{kalesaki2013,delerue2014,delerue2015}. A
strong quantum mechanical coupling induces an important wavefunction overlap between nanocrystals and therefore large
STS resonances. Inversely, a weak coupling give narrow resonance widths. The average full-widths of the e$_{\rm{1}}$ and
h$_{\rm{1}}$ resonances of the superlattices are 103 meV $\pm$ 5 meV (e$_{\rm{1}}$) and 110 $\pm$ 20 meV respectively,
see figures \ref{fig:figure3}(a) and (b). From our atomistic tight-binding calculations, it is possible to extract the
full-width dispersion of the different band folds by taking the difference in energy between the highest and lowest
energy bands in each fold.  The result is presented in figure \ref{fig:figure2}(d). As expected, the band width
increases with increasing $d$ for both the e$_{\rm{1}}$ and h$_{\rm{1}}$ bands, for both nanocrystal sizes considered.
The experimental values (see figure \ref{fig:figure3}) are very close to our tight-binding calculations for $d$/$D$ =
0.5. In addition to an atomistic tight-binding model, the electronic structure of the superlattice can also be modeled
using a tight-binding approach using the S$_{\rm{e}}$ and S$_{\rm{h}}$ envelope states of the QDs themselves. To first
order, the dispersion of the e$_{\rm{1}}$ and h$_{\rm{1}}$ bands is proportional to 2t$_{\rm{1}}$ and 2t$_{\rm{2}}$,
where t$_{\rm{1}}$ and t$_{\rm{2}}$ describe the effective coupling strength between the nanocrystals due to
S$_{\rm{e}}$ (e$_{\rm{1}}$) and S$_{\rm{h}}$ (h$_{\rm{1}}$) states.  Hence, we find an upper limit for the electronic
coupling between nanocrystals  of 60 meV and 65 meV for the e$_{\rm{1}}$ and h$_{\rm{1}}$ states, respectively.

Fluctuations in the width of the peaks can arise due to differences in the necking (and therefore coupling strength, see
figures \ref{fig:figure1}(a) and \ref{fig:figure3}). We observe significant fluctuations in the width of the resonances
(and couplings) for both electron and hole states. This is consistent with the connectivity of nanocrystals observed in
TEM.  In addition, the magnitude of the intervalley coupling, and therefore the peak width, depends on QDs
size\cite{overgaag2009}. Coupling of the tunnelling electrons with acoustic or optical phonons can increase the peak
width. At 5 K, the temperature of the measurement, extra phonon resonances are expected at the high-energy side of the
principal peak.  This has been observed in the case of CdSe nanocrystals\cite{sun2009}.  However asymmetric peak
broadening is not observed in the present case, excluding electron-phonon coupling as a main physical origin of
broadening.  It is of interest to compare the present results with those obtained on individual QDs and for a lattice
with close-packed PbSe NC ordering that has been strongly annealed\cite{liljeroth2005}. In the former, the width of the
resonances are smaller (around 75 $\pm$ 15 meV) than the values found in our measurements\cite{overgaag2009}.
In the latter, the variations in the widths of the resonances and the band gaps were much more pronounced than observed
here.  In the present case, the band structure of the square superlattices and the width of the resonances are much more
uniform, with a moderate electronic coupling between the NCs.

The TEM and STM images show that there is a distribution in the size of the building blocks and the width of the
crystalline necks between the nanocrystals (see figure \ref{fig:figure1}). A certain variation in the energy of the
on-site $s$- and $p$-states should therefore be expected. However, our results indicate that the variations of the
on-site energies are small (see S.I). Hence, we attribute the fluctuations in peak width to differences in the width and
crystallinity of the necks between individual QDs.

Careful inspection of the dI/dV spectra in the gap region reveals that there may be some in-gap states present in the
array. It is unlikely to be noise because the amplitude of those peaks saturates with high current set-point (see S.I.).
The presence of those peaks are attributed to trap states. Such trap states, could come from the detachment of ligands
from certain selenium and lead atoms at the surface of the nanocrystals as seen for other II-VI
nanocrystals\cite{houtepen2017}. The saturation is explained by state filling for these localized states, while
tunneling via the delocalized band states remains in the shell-tunneling regime\cite{berthe2008}. Those states are
highly localized states at the surface of the nanocrystals, causing a drastic reduction of the probability for electrons
in those states to tunnel out of the nanocrystal. The shell-filling regime is often associated with double peaks due to
the additional energy needed to inject another electron in those already occupied states. Here, no double peak is
visible because of the large dielectric constant of the PbSe resulting in Coulomb energy of only a few meV. 

Superlattices of semiconductors nanocrystals are considered to become a class of novel electronic materials once the
electronic coupling between the NCs is strong enough, and once the strucral disorder is sufficiently small. There is an
ongoing discussion on the nature of the carrier transport and convincing evidence for band-like transport has not been
presented yet.

In order to improve the electronic coupling and lower the disorder. It is imperative to use a
nanocrystal solution with the smallest possible distributions in size and shape. The shape of nanocrystals can be
modified by the ligand density, leading to smoother nanocrystalline facets\cite{peters2017}. The forces that keep the
nanocrystals in the assembly plane can be tweaked by having higher solvent-air surface tension\cite{soligno2019}. One of
the sources of non-uniform necking is the angular variation between [100]
facets\cite{peters2018,walravens2019,mccray2019} and, assembly and post-treatment techniques aim to improve the
regularity of such superlattices.

\section*{Conclusion}

In conclusion, we have measured the local density of states of a square superlattice made of PbSe semiconductor
nanocrystals. The STM image of the annealed sample revealed a clean surface where it was possible to acquire
reproducible scanning tunneling spectra. Those spectra showed a band gap of 0.66 eV in agreement with tight-binding
calculations and optical spectroscopy measurements. The peak widths for the first energy levels were close to our
atomistic tight-binding calculations for coupling of $d$/$D$ = 0.5. By comparing the experimentally determined
bandwidths with tight-binding simulations, we find an upper limit of 65 meV for the coupling (hopping term) between the
nanocrystals in the superlattice. There are significant fluctuations in the observed band-widths, consistent with large
differences in the necking between QDs as observed in TEM images. Band-like electronic coupling will require more
homogeneous couplings and ideally stronger crystalline neckings between the nanocrystals.

\section*{Acknowledgement}
D.V. wishes to acknowledge the European Research Council for his Advanced Grant FIRST STEP, 692691

\section*{References}

\bibliographystyle{iopart-num}
\bibliography{bibliography.bib}

\end{document}